# Diffusion theory for the infection pathway of virus in a living cell

Yuichi Itto

*Science Division, Center for General Education, Aichi Institute of Technology, Aichi 470-0392, Japan*

**Abstract.** The infection pathway of virus in living cell is of interest from the viewpoint of the physics of diffusion. Here, recent developments about a diffusion theory for the infection pathway of an adeno-associated virus in cytoplasm of a living HeLa cell are reported. Generalizing fractional kinetics successfully modeling anomalous diffusion, a theory for describing the infection pathway of the virus over the cytoplasm is presented. The statistical property of the fluctuations of the anomalous-diffusion exponent is also discussed based on a maximum-entropy-principle approach. In addition, an issue regarding the continuum limit of the entropy introduced in the approach is carefully examined. The theory is found to imply that the motion of the virus may obey a scaling law.

# 1. INTRODUCTION

Viruses and related phenomena are of great interest from the viewpoint of physics (see, for example, Refs. [1,2]). In particular, understanding the infection pathway of virus in living cell may be relevant, for example, to drug delivery based on virus-based carriers [3].

The purpose of the present article is to report recent developments about a diffusion theory for the infection pathway of an adeno-associated virus in cytoplasm of a living HeLa cell. It is discussed that the exponent characterizing the diffusion property of the virus fluctuates depending on localized areas of the cytoplasm. Then, there is no information on the local property of such fluctuations. Therefore, as will be seen later, the entropy associated with the local fluctuations is introduced in the theory: a measure of uncertainty about how the exponent locally distributes over the cytoplasm. It turns out to play a key role for proposing the statistical property of the fluctuations, i.e., the statistical distribution of the fluctuations over the cytoplasm, which is crucial in the theory. In addition, the discussion about the entropy will be developed further (see Sec. 3 below).

In the experiments in Refs. [4,5], it has been observed, by using the technique of real-time single-molecule imaging, that the viruses, each of which is labeled with a fluorescent dye molecule, exhibit stochastic motions in the cytoplasm in both the form being confined in the endosome, i.e., a spherical vesicle and the non-confined form. Based on analysis of the trajectory of the virus, then the mean square displacement, which is denoted here by $\overline{x^2}$ with the over-bar being the average, has been evaluated in order to characterize the diffusion property of the virus. For large elapsed time, $t$, $\overline{x^2}$ has been found to behave as follows:

$$\overline{x^2} \sim t^{\alpha}. \qquad (1)$$

The resulting diffusion property has two cases: one is normal diffusion with $\alpha = 1$, and the other is the case with $0 < \alpha < 1$, which is referred to as anomalous diffusion. Here is a certainly remarkable feature [4] that the exponent, $\alpha$, in the case of anomalous diffusion fluctuates between $0.5$ and $0.9$, depending on localized areas of the cytoplasm. Therefore, this is apparently different from traditional anomalous diffusion [6,7] widely discussed in the literature.

It may be of interest to mention that the variation of the exponent has been observed for diffusion of macromolecule (such as ribosome) in a bacterial cytoplasm in a recent work [8]. There, nonspecific interactions, which are due to high macromolecular concentration, have been discussed for such a variation. (Later, we will briefly discuss a possible relevance of nonspecific interactions to the origin of anomalous diffusion of the virus.)

## 2. FRACTIONAL KINETICS AND ITS GENERALIZATION

Consider 1-dimensional stochastic motion of the virus in the cytoplasm. As a first step, we regard the cytoplasm as a medium for stochastic motions of the virus in both the endosomal and non-confined forms. This medium is then imaginarily divided into many small blocks, each of which is identified with a localized area of the cytoplasm. To describe the motion of the virus in a given block, we apply fractional kinetics [9] modeling anomalous diffusion in a unified way. The fractional diffusion equation we consider here is as follows:

$$\frac{\partial f(x,t)}{\partial t} = {}_0\mathcal{D}_t^{1-\alpha} D_\alpha \frac{\partial^2 f(x,t)}{\partial x^2}. \qquad (2)$$

Here, $f(x,t)\,dx$ is the probability of finding the virus in the interval $[x, x+dx]$ at time $t$, $D_\alpha$ is a generalized diffusion constant, and ${}_0\mathcal{D}_t^{1-\alpha}$ is a fractional operator [9,10] defined by ${}_0\mathcal{D}_t^{1-\alpha} = (\partial/\partial t)\,{}_0\mathcal{D}_t^{-\alpha}$ with ${}_0\mathcal{D}_t^{-\alpha} g(t) = [1/\Gamma(\alpha)] \int_0^t dt'(t-t')^{\alpha-1} g(t')$, where $\Gamma(\alpha)$ is the

Euler gamma function. $\alpha$ in Eq. (2) is taken to be in the following range:

$$0 < \alpha < 1. \tag{3}$$

The scheme of continuous-time random walks [11] tells us the physical basis behind Eq. (2). As discussed in Ref. [12], two different distributions are relevant in this scheme: one for a spatial displacement $\Delta$ of the virus in a finite time step $\tau$, and the other for the time step $\tau$. The former is sharply peaked at $\Delta = 0$ and has evenness with respect to $\Delta$, whereas the latter is implied to decay as $\sim s^\alpha / \tau^{1+\alpha}$ for long time step, i.e., a power law characterized by $\alpha$, which has the divergent first moment. Here, $s$ is supposed to be a characteristic time, at which the virus is displaced. It can be found [12] that $D_\alpha$ is given by $D_\alpha = \langle \Delta^2 \rangle / (2s^\alpha)$ with $\langle \Delta^2 \rangle$ being the second moment of $\Delta$.

With the initial condition, $f(x, 0) = \delta(x)$, the mean square displacement of the virus turns out to have the form in Eq. (1). Thus, the behavior observed in a localized area of the cytoplasm is reproduced. Note that normal diffusion is realized in the limit, $\alpha \to 1$, in our present theory.

Next, let us generalize fractional kinetics mentioned above in order to describe the motion of the virus over the cytoplasm. In Ref. [12], such a discussion has been made by introducing the statistical fluctuation of $\alpha$. There, a basic premise is the existence of two largely separated time scales in the infection pathway: the time scale of variation of exponent fluctuations is much larger than that of stochastic motion of the virus in each local block. In other words, $\alpha$ in each local block slowly varies in time, but is assumed to be approximately constant. Denoting the statistical distribution of the fluctuations of $\alpha$ by $P(\alpha)$, the following generalized fractional diffusion equation has been presented:

$$\int d\alpha\, P(\alpha)\, s^{\alpha-1}\, {}_0\mathcal{D}_t^{-(1-\alpha)} \frac{\partial f(x,t)}{\partial t} = D \frac{\partial^2 f(x,t)}{\partial x^2}, \tag{4}$$

where $D$ is the diffusion constant given by $D = \langle \Delta^2 \rangle / (2s)$. We here mention the following recent study. In Ref. [13] (see also Ref. [14]), a theoretical framework has been developed for deriving Eq. (4). In this framework, in contrast to the procedure employed in Ref. [12], the existence of the large time-scale separation is *explicitly* taken into account. In the above-mentioned scheme, it turns out [13] that Eq. (4) appears through the average of the distribution of time step with respect to $P(\alpha)$. For the virus in a given local block, it can be shown [12,13] that taking $P(\alpha) = \delta(\alpha - \alpha_0)$ with $\alpha_0$ being a certain fixed exponent in the range $0 < \alpha_0 < 1$ and applying the operator $_0\mathcal{D}_t^{1-\alpha_0}$, Eq. (4) becomes reduced to Eq. (2). Thus, fractional kinetics is generalized in this way.

## 3. STATISTICAL DISTRIBUTION OF EXPONENT FLUCTUATIONS AND MAXIMUM-ENTROPY-PRINCIPLE APPROACH

Clearly, it is necessary to determine the statistical distribution $P(\alpha)$, since otherwise Eq. (4) is formal. Here, we wish to present a proposition for it based on both the experimental data and a maximum-entropy-principle approach. In addition, we carefully examine an issue concerning with the continuum limit of the entropy in the approach, which has not been discussed in Refs. [12-14].

For 104 trajectories of the viruses, the mean square displacement has the form in Eq. (1) [4]: 53 trajectories among them exhibit normal diffusion, whereas other 51 show anomalous diffusion with the exponent $\alpha$ varying between $0.5$ and $0.9$. (Although there are trajectories yielding a parabolic form [4,5], such trajectories have been neglected since the number of them is seen to be less compared to those in the case of normal diffusion and anomalous diffusion.) In addition, the virus tends to reach the nucleus of the cell. Accordingly, these facts motivated the works in Refs. [12,13] to suggest that normal diffusion is often to be realized, whereas anomalous diffusion with the exponent near $\alpha = 0$ may not be the case.

There, it is also supposed that the exponent found in the endosomal form is slightly different from that found in the non-confined form. Based on these considerations, the following exponential distribution of the fluctuations has been proposed:

$$\hat{P}(\alpha) \propto e^{\lambda \alpha}, \tag{5}$$

where $\lambda$ is a positive constant.

Now, as will be seen below, it is possible to theoretically derive the distribution in Eq. (5) in a consistent manner. In Refs. [12,13], the medium is viewed as a collection formed by constructing the local blocks. This construction then offers all of possible distinct collections in the sense that each collection is different from each other in terms of the local fluctuations and no difference exists at the statistical level of the fluctuations. That is, the local property of exponent fluctuations is distinct depending on the collections, but the statistical property is not. The local blocks seem to be independent in terms of the exponent. For a set of discrete values of different exponents, $\{\alpha_i\}_i$, the total number of distinct collections, $G$, is therefore calculated as a combinational problem and is given by

$$G = \frac{N!}{\prod_i n_{\alpha_i}!}, \tag{6}$$

where $N$ is the total number of blocks in the medium, $n_{\alpha_i}$ is the number of blocks with the $i$ th value of the exponent, $\alpha_i$, in the medium, and $\sum_i n_{\alpha_i} = N$ is fulfilled. Then, the entropy associated with the local fluctuations has been introduced as follows:

$$S = \frac{\ln G}{N}. \tag{7}$$

As mentioned in the INTRODUCTION, this gives a measure of uncertainty about the local property of the fluctuations over the medium. Here, $N$ and $n_{\alpha_i}$'s are assumed to be large, since the medium is composed of many blocks. In Ref. [15], it has been shown that this assumption seems to be appropriate. It is found [12,13] that the entropy in Eq. (7) can approximately be given by the form of the Shannon entropy:

$$S \cong -\sum_i P_{\alpha_i} \ln P_{\alpha_i}, \tag{8}$$

where $P_{\alpha_i} = n_{\alpha_i} / N$ is the probability of finding the exponent $\alpha_i$ in a given local block of the medium. Correspondingly, the entropy in the case of continuous values is taken to be

$$S[P] = -\int d\alpha \, P(\alpha) \ln P(\alpha), \tag{9}$$

where $P(\alpha) d\alpha$ is the probability of finding the exponent in the interval $[\alpha, \alpha + d\alpha]$.

Now, a careful treatment should be employed for the continuum limit of the entropy in Eq. (8). Below, following the discussion in Ref. [16], we shall see it in the present context.

Let us divide a fixed interval $[a, b]$ into a set of intervals, $\{[\alpha_i, \alpha_{i+1}]\}_{i=1,2,\ldots,n}$, as $a = \alpha_1 < \alpha_2 < \cdots < \alpha_{n+1} = b$. Then, we increase the number of discrete exponents $\alpha_i$'s according to some definite function of $\alpha$ denoted by $m(\alpha)$, which determines how each interval tends to zero in the limit $n \to \infty$. In this limiting procedure, the following relation holds [16]:

$$\lim_{n \to \infty} n(\alpha_{i+1} - \alpha_i) = \frac{1}{m(\alpha_i)}. \tag{10}$$

In other words, the measure, $m$, is introduced in this way. At this stage, the limiting procedure is generically treated since the form of $m$ may depend on the medium, but, as will be seen below, $m$ is taken to be a certain constant. This measure has the following property: $\lim_{n \to \infty} \sum_{i=1}^{n} [nm(\alpha_i)]^{-1} = \int_a^b d\alpha.$ The probability $P_{\alpha_i}$ in Eq. (8) is connected to the probability density $P(\alpha_i)$ in Eq. (9) through the measure as follows:

$$P_{\alpha_i} = P(\alpha_i)(\alpha_{i+1} - \alpha_i)$$
$$\to P(\alpha_i) \frac{1}{nm(\alpha_i)}. \tag{11}$$

Taking into account the above property, the entropy in Eq. (8) tends to

$$S \to -\int_a^b d\alpha\, P(\alpha) \ln \frac{P(\alpha)}{nm(\alpha)}. \tag{12}$$

We here note the following points. Due to the presence of the additive logarithmic divergence, $\lim_{n \to \infty} \ln n$, the quantity in Eq. (12) itself fails to define the entropy for a continuous variable. This difficulty is, however, overcome when the change of the entropy, not the absolute value of the entropy, is considered, leading to cancellation of such a divergence. Then, the form of $m(\alpha)$ may depend on the medium as mentioned above, but there is no *a priori* information about determination of the form, and this is precisely the situation we are considering here. In such a situation, it seems natural to suppose that all of possible exponents with equal intervals should be taken into account in the fixed interval. In other words, it seems fair to assume that any intervals in the above-mentioned set $\{[\alpha_i, \alpha_{i+1}]\}_{i=1,2,...,n}$ equally tend to zero each other in the limit $n \to \infty$, implying that $m(\alpha)$ can be taken as a certain constant. This turns out to bring an additive constant in Eq. (12), which is cancelled again in the entropy change.

Thus, from these considerations, we can define $S[P]$ in Eq. (9) as the entropy in the case of continuous values of the exponent.

We shall show that the distribution in Eq. (5) can be derived based on maximization of the Shannon entropy in Eq. (9). We are considering the situation that only information is available about the statistical property of exponent fluctuations. In such a situation, we impose two constraints: one for the normalization condition, $\int d\alpha P(\alpha) = 1$, and the other for the expectation value of $\alpha$, $\int d\alpha P(\alpha)\alpha = \overline{\alpha}$. Under these constraints, we maximize $S[P]$ with respect to $P(\alpha)$ as follows:

$$\delta_P \left\{ S[P] - \kappa \left( \int d\alpha P(\alpha) - 1 \right) + \lambda \left( \int d\alpha P(\alpha)\alpha - \overline{\alpha} \right) \right\} = 0, \quad (13)$$

where $\kappa$ and $\lambda$ are, respectively, the Lagrange multipliers associated with the constraints on the normalization condition and the expectation value, and $\delta_P$ denotes the variation with respect to $P(\alpha)$. It should be noted that the condition, $P(1) > P(0)$, has been imposed in Eq. (13), which requires $\lambda$ to be a positive Lagrange multiplier. This condition is supposed to express the tendency that the virus reaches the nucleus. The stationary solution of Eq. (13) is given by $\check{P}(\alpha) \propto e^{\lambda \alpha}$, which is, in fact, the distribution in Eq. (5).

In the above derivation, we have imposed the constraints on the expectation value of $\alpha$ as well as the normalization condition. If additional information on the expectation values of some relevant quantities, $\overline{Q^{(k)}(\alpha)} = \int d\alpha P(\alpha) Q^{(k)}(\alpha)$, where $Q^{(k)}(\alpha)$ is the $k$ th quantity, is given, then the maximum-entropy-principle approach reads

$$\delta_P \left\{ S[P] - \kappa \left( \int d\alpha P(\alpha) - 1 \right) - \lambda' \left( \int d\alpha P(\alpha) \alpha - \overline{\alpha} \right) \right.$$
$$\left. - \sum_k \lambda^{(k)} \left( \int d\alpha P(\alpha) Q^{(k)}(\alpha) - \overline{Q^{(k)}(\alpha)} \right) \right\} = 0, \quad (14)$$

where $\lambda'$ and $\lambda^{(k)}$'s are, respectively, the Lagrange multipliers associated with the constraints on the expectation values of $\alpha$ and $Q^{(k)}(\alpha)$'s. The stationary solution derived above is changed into

$$\breve{P}(\alpha) \propto \exp[-\lambda'\alpha - \sum_k \lambda^{(k)} Q^{(k)}(\alpha)], \quad (15)$$

accordingly. The multipliers appearing here are required to satisfy a relation to be suggested by the above condition $P(1) > P(0)$. Thus, the present approach also enables one to examine $\breve{P}(\alpha)$ for describing the statistical fluctuation to be observed in the experiment, if such a fluctuation distribution is different from Eq. (5).

Closing this section, we briefly discuss a possible relevance of nonspecific interactions to the origin of anomalous diffusion of the virus. It has been considered [4,5] that such an origin is due to the presence of obstacles (e.g., organelles) in the cytoplasm. Now, numerical simulation combined with experimental data has been performed in Ref. [8] for modeling diffusion of macromolecules in a bacterial cytoplasm. There, it has been discussed that the bacterial cytoplasm is ploy-disperse with high concentration of macromolecules and interactions between them are repulsive with and without nonspecific attraction, demonstrating that macromolecules exhibit both normal diffusion and anomalous diffusion. In particular, it has been shown that diffusion is suppressed in the case when nonspecific attraction is taken into account (see, for example, Ref. [17] for a similar discussion). Therefore, from the above-mentioned consideration, this may imply in our present case that such nonspecific interactions between the virus and obstacles can lead to anomalous diffusion of the virus.

## 4. SCALING LAW FOR THE MOTION OF THE VIRUS

Let us discuss the motion of the virus over the cytoplasm based on Eq. (4) with the distribution in Eq. (5). As shown in Ref. [13], it is implied that $f(x,t)$ asymptotically behaves for large elapsed time as follows:

$$f(x,t) \sim \frac{1}{\sqrt{4\langle\Delta^2\rangle_\lambda \ln[t/(e^\lambda s)]}} \exp\left(-\frac{|x|}{\sqrt{\langle\Delta^2\rangle_\lambda \ln[t/(e^\lambda s)]}}\right), \qquad (16)$$

where $\langle\Delta^2\rangle_\lambda$ is defined by $\langle\Delta^2\rangle_\lambda = [(e^\lambda-1)/(2\lambda)]\langle\Delta^2\rangle$. Then, we immediately see that it satisfies the following scaling law:

$$f(x,t) \sim \frac{1}{\sqrt{\ln[t/(e^\lambda s)]}} \hat{f}\left(\frac{x}{\sqrt{\ln[t/(e^\lambda s)]}}\right), \qquad (17)$$

where $\hat{f}(x)$ is a scaling function defined by $\hat{f}(x) = (1/\sqrt{4\langle\Delta^2\rangle_\lambda})\exp(-|x|/\sqrt{\langle\Delta^2\rangle_\lambda})$. Accordingly, the spatial extension of $f(x,t)$, e.g., its half-width, $l$, is seen to be

$$l \sim \sqrt{\ln\frac{t}{e^\lambda s}}, \qquad (18)$$

which indicates that the motion of the virus may exhibit logarithmic behavior.

We also mention the following. In Ref. [12], the mean square displacement of the virus has been calculated based on Eqs. (4) and (5). From it, the root-mean square displacement, $\sqrt{\overline{x^2}}$, turns out to have the form in Eq. (18). Therefore, one might think that the difference between $l$ and $\sqrt{\overline{x^2}}$ is nothing but the characterization of the logarithmic behavior. Regarding this point, we emphasize the following fact. The behavior of $l$ comes from $f(x,t)$ in Eq. (16) [or, equivalently Eq. (17)] itself, and accordingly, is seen to be more fundamental than that of $\sqrt{\overline{x^2}}$, which is based on the average with respect to $f(x,t)$ in Eq. (4) [with the distribution in Eq. (5)].

It is of extreme interest to further examine the infection pathway of the virus over the cytoplasm: if the scaling law in Eq. (17) can experimentally be observed, then the time-scale separation is expected to exist in the infection pathway.

**CONCLUSION**

We have reported recent developments about a diffusion theory for the infection pathway of an adeno-associated virus in cytoplasm of a living HeLa cell. The generalized fractional kinetics has been discussed for describing the infection pathway of the virus over the cytoplasm. A proposition for the statistical distribution of exponent fluctuations has been presented. The entropy associated with the fluctuations has been introduced, and the discussion about a careful treatment for the continuum limit of the entropy has been developed. Then, we have seen that maximization condition of the entropy leads to the proposed distribution. We have also mentioned a scaling nature of the motion of the virus.

As mentioned in this article, the exponent slowly varies but is assumed to be approximately constant. If this assumption is relaxed, then the statistical distribution may deviate from Eq. (5), in general. This leads to the following question: is it possible to determine the behavior of such a deviation? In Ref. [15], this question is affirmatively answered for a class of small deviations. There, it is shown that the deviation obeys the multivariate Gaussian distribution. Therefore, it is also of extreme interest to examine if such a deviation can experimentally be observed in the infection pathway.

**CONFLICT OF INTEREST**

Declared none.

**ACKNOWLEDGEMENTS**

The present article is based on the author's talk at the Drug Discovery and Therapy World Congress 2016 (22-25 August, 2016, Boston, USA). The author would like to thank the organizers of the congress for the invitation. He also acknowledges the Aichi Institute of Technology for support.

*Note added*. Quite recently, the maximum-entropy-principle approach has been applied for diffusion of virus capsid in a different cell in Ref. [18]. There, a statistical distribution of exponent fluctuations has been derived in accordance with the one observed in a relevant experiment.